\documentstyle{aipproc2}

\include{psfig}

\begin{document}

\newcommand{\R}{{\mathbb R}}
\newcommand{\C}{{\mathbb C}}
\newcommand{\Z}{{\mathbb Z}}
\newcommand{\be}{\begin{equation}}
\newcommand{\ee}{\end{equation}}
\newcommand{\bea}{\begin{eqnarray*}}
\newcommand{\eea}{\end{eqnarray*}}
\newcommand{\tb}{\tilde{\beta}}
\newcommand{\ta}{\tilde{a}}
\newcommand{\tal}{\tilde{\alpha}}
\newcommand{\ttau}{\tilde{\tau}}
\newcommand{\td}{\tilde{\delta}}
\newcommand{\cQ}{{\mathcal{Q}}}
\newcommand{\tQ}{{\tilde{\cQ}}}
\newcommand{\cE}{{\mathcal{E}}}
\newcommand{\cF}{{\mathcal{F}}}
\newcommand{\cP}{{\mathcal{P}}}
\newcommand{\tG}{{\tilde{\cG}}}
\newcommand{\tN}{{\tilde{N}}}
\newcommand{\tF}{{\tilde{F}}}
\newcommand{\tV}{{\tilde{V}}} 
\newcommand{\cH}{{\mathcal{H}}}
\newcommand{\cX}{{\mathcal{X}}}  
\newcommand{\cY}{{\mathcal{Y}}}
\newcommand{\cR}{{\mathcal{R}}}
\newcommand{\cT}{{\mathcal{T}}}
\newcommand{\cA}{{\mathcal{A}}}  
\newcommand{\lt}{\frac{\Lambda}{3}}
\newcommand{\cM}{{\mathcal{M}}}
\newcommand{\tE}{\tilde{E}}
\newcommand{\cp}{{\mathcal{p}}}
\newcommand{\tPhi}{\tilde{\Phi}}

\newcommand{\nn}{\nonumber}
\newcommand{\tA}{\tilde{A}}
\newcommand{\tp}{\tilde{\phi}}
\newcommand{\e}{\varepsilon}
\newcommand{\mbin}{\mbox{in}}
\newcommand{\mbout}{\mbox{out}}

\newcommand{\az}{\alpha}
\newcommand{\bz}{\beta}
\newcommand{\cz}{\gamma}
\newcommand{\gz}{\gamma}
\newcommand{\dz}{\delta}
\newcommand{\ez}{\epsilon}
\newcommand{\kz}{\kappa}
\newcommand{\lz}{\lambda}
\newcommand{\na}{\nabla}

\newcommand{\ou}{\tilde{u}}
\newcommand{\on}{\tilde{n}}
\newcommand{\oN}{\tilde{N}}
\newcommand{\oV}{\tilde{V}}
\newcommand{\ok}{\tilde{k}}
\newcommand{\oa}{\tilde{a}}
\newcommand{\oep}{\tilde{\varepsilon}}
\newcommand{\oj}{\tilde{\jmath}}
\newcommand{\os}{\tilde{s}}

\newcommand{\bT}{\underline{T}}
\newcommand{\bI}{\underline{I}}

\title{Quasilocal energy and naked black holes}

\author{Ivan Booth and Robert Mann}

\address{Department of Physics,University of Waterloo, Waterloo, Ontario, N2L 3G1}

\maketitle

\begin{abstract}

We extend the Brown and York notion of quasilocal energy to
include coupled electromagnetic and dilaton fields and also 
allow for spatial boundaries that are not orthogonal to the 
foliation of the spacetime. We investigate how the quasilocal
quantities measured by sets of observers transform with respect to
boosts. 
As a natural application of this work we investigate the naked 
black holes of Horowitz and Ross calculating the quasilocal 
energies measured by static versus infalling observers.

\end{abstract}

\section{Introduction}

The definition of energy in general relativity continues to be
an area of active research. It is widely accepted that 
while one cannot localize gravitational energy and therefore define
a gravitational stress energy tensor one can define a notion
of the total energy in a spacetime (for example the ADM or Bondi
masses). In between these two extremes one can define energy
quasilocally -- that is define the amount of energy contained
in a finite volume of spacetime.

One popular definition of quasilocal energy (QLE)
was proposed by
Brown and York in 1993 \cite{BY}. As we shall see in the next 
section their approach derives a Hamiltonian from the standard
Einstein-Hilbert action and a notion of quasilocal energy from 
that Hamiltonian. Since its proposal this quasilocal energy
has found application in gravitational thermodynamics
and the study of the production of pairs of black holes.
It has been shown to reduce to the ADM and Bondi masses in the
appropriate limits as well as a Newtonian notion of gravitational
energy for certain specific examples. References for these
may be found in \cite{naked}. 

In this paper, we extend this notion of quasilocal energy to
include coupled electromagnetic and dilaton fields and also 
allow for spatial boundaries that are not orthogonal to the 
foliation of the spacetime. Using the second generalization 
we can calculate the quasilocal energies measured by 
sets of observers who are moving around in a spacetime.
We see that the QLE transforms in a 
Lorentzian way under boosts of the observers.
In the last section we find a natural application for this work in
the naked black holes first studied by Horowitz and Ross \cite{HR}.
Such black holes have small curvature constants yet observers
falling into them experience massive tidal forces as they approach 
the event horizon. We calculate the energies
measured by these observers.

\section{Quasilocal Energy}

In classical mechanics the action $I$ of a point particle is the time
integral
of its kinetic energy minus its potential energy. To wit, 
\bea
I = \int dt (p \dot{q} - H) 
\eea
where $p$ is the particle momentum, $q$ is its position, and $H$ is its
Hamiltonian/potential energy. Taking the first variation of this action we
obtain the equations of motion of the particle (plus certain boundary
conditions that must be satisfied so that the variation will vanish). 

\begin{figure}
\vspace{6cm}
\centerline{\psfig{figure=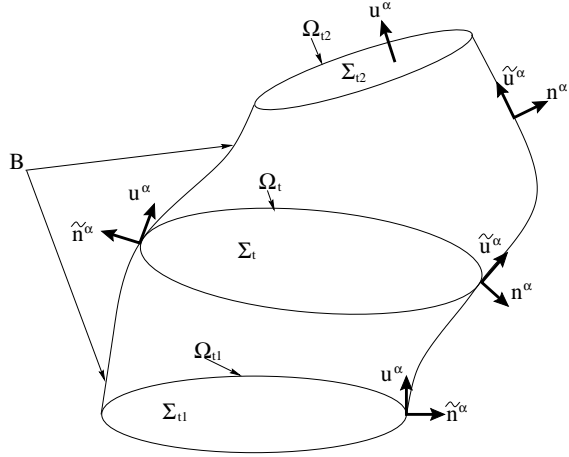,height=6cm,angle=270}} 
\caption{A three dimensional schematic of the region $M$, 
its assorted normal vector fields, and a 
typical element of the foliation.}
\label{fig} 
\end{figure}

Now the action for gravity is well known so it is natural to extend and
reverse this procedure to define a Hamiltonian for gravity. We follow
the procedure of Brown and York \cite{BY}. Given a region of spacetime $M$
(figure \ref{fig}) the role of time is played by a foliation
$\Sigma_t$ of
the region and an accompanying 
timelike vector field $T^\az$ such that $T^\az
\partial_\az t = 1$. With respect to the foliation we can write 
this vector
field in terms of a lapse function $N$ and spacelike 
shift vector field $V^\az$. Namely, 
$T^\az = N u^\az + V^\az$
where $u^\az$ is the unit normal to $\Sigma_t$.
The region is finite and  bounded by a timelike three surface $B$ with
unit normal $n^\az$ and two spacelike surfaces $\Sigma_1$ 
and $\Sigma_2$.
$\Sigma_t$ induces a foliation $\Omega_t$ of $B$ which has normals
$\tilde{u}^\az$ and $\tilde{n}^\az$ when viewed as embedded in $B$ and
$\Sigma_t$ respectively. Brown and York assumed $u^\az = \tilde{u}^\az$
and $n^\az = \tilde{n}^\az$ (equivalently $\eta = n^\az u_\az = 0$). 
We drop that assumption here and allow for nonorthogonal foliations of
$M$ (as we did in \cite{NO}). 

Define $K_{\az \bz}$ and $\Theta_{\az \bz}$ as the extrinsic curvatures of
the surfaces $\Sigma_t$ and $B$ in $M$ respectively. $k_{\az \bz}$
is the extrinsic curvature of $\Omega_t$ in $\Sigma_t$. $g_{\az \bz}$,
$\gamma_{\az \bz}$, $h_{\az \bz}$, and $\sigma_{\az \bz}$ are the metrics on $M$, $B$, $\Sigma_t$, and $\Omega_t$ respectively. 
We assume that the boundary $B$ is generated by Lie dragging an $\Omega_t$ along $T^\az$. Physically we view $B$ as being the 
history of a set of observers evolving according to the 
vector field $T^\az$. 
 
Allowing electromagnetic $F_{\az \bz}$ and dilaton $\phi$ fields
(whose coupling if governed by the constant $a$) the regular 
Hilbert action for general relativity over the region $M$ is
\begin{eqnarray}
I &=& \frac{1}{2 \kappa} \int_M d^4 x \sqrt{-g} 
\left(\cR - 2 \Lambda  
- 2 (\nabla_\az \phi)(\nabla^\az \phi) - e^{-2a\phi}
 F_{\az \bz} F^{\az \bz} \right) \nn \\ 
&& + \frac{1}{\kappa} \int_\Sigma d^3 x \sqrt{h} K -
\frac{1}{\kappa} \int_B d^3 x \sqrt{- \gamma} \Theta \nn +
\frac{1}{\kappa}
\int_\Omega d^2 x \sqrt{\sigma} \sinh^{-1} (\eta)
 - \bI. \nn 
\end{eqnarray}
$\kappa = 8 \pi$ in units where $G = c = 1$.
The boundary terms ensure that if we keep metrics and certain
matter terms fixed on the boundaries then
the variational principle is properly defined. The $\underline{I}$ term is
any functional defined with respect to the boundary metric $\gamma_{\az
\bz}$. Since that metric is kept constant during the variation, $\delta
\underline{I} = 0$ and so doesn't affect the equations of motion. This
term is called the reference term and as we shall see a bit later,
its form determines the zero of the action. 
 
We break up this action with respect to the foliation to obtain
\begin{eqnarray}
I &=& \int dt \int_{\Sigma_t} d^3 x \left\{ P_h^{\az \bz} 
\pounds_T h_{\az \bz} + P_\phi \pounds_T \phi + P_{ \bar{A}}^\az 
\pounds_T ( \bar{A}_\az) \right\}  
+ \int dt \int_{\Omega_t} d^2 x \left\{ P_{\sqrt{\sigma}}
(\pounds_T \sqrt{\sigma}) \right\} \nn \\
&& - \int dt \int_{\Sigma_t} d^3 x \left\{ N \cH^m + V^\az \cH^m_\az + 
(T^\az A_\az) \cQ \right\}
- \int dt \int_{\Omega_t} d^2 x \sqrt{\sigma} 
\left\{ \oN (\oep + \oep^m)  - \oV^\az (\oj_\az + \oj_\az^m) 
\right\} - \underline{I}.
\nn
\end{eqnarray}
The $P$'s are momentum terms, $\pounds_T$ is the Lie derivative in the
direction $T$. $\cH^m$ and $\cH_\az^m$ are the regular
gravitational Hamiltonian constraint equations and $\cQ$ is an
electromagnetic constraint. $A_\az$ is the electromagnetic vector
potential, $\bar{A}_\az$ is the projection of that potential into
$\Sigma_t$, and $\tilde{N}$ and $\tilde{V}^\az$ are the lapse 
and shift for the
foliation of the boundary $B$. Then by analogy with the point particle
action, we can define a Hamiltonian
\bea
H^m = \int_{\Omega_t} d^2 x \sqrt{\sigma} \left\{ \tN (\oep + \oep^m)
- \oV^\az (\oj_\az + \oj^m_\az) \right\} - \underline{H}, \nn
\eea
where we have assumed that the constraint equations are satisfied. 
The $\oep$ terms are QLE densities and the $\oj_\az$ terms
are angular momentum densities. $\underline{H}$ is the reference term
calculated from $\underline{I}$. We define the 
quasilocal energy
to be the Hamiltonian for observers measuring proper time
($\tilde{N}=1$) who are at rest with respect to the leaf
$\Omega_t$ ($\oV^\az = 0$). 

\subsection{Calculating the quasilocal energy}

We now consider the exact form of the quasilocal energy densities. 
\bea
\oep = \frac{1}{8 \pi} \tilde{k} = - \frac{1}{8 \pi} \sigma^{\az \bz}
\nabla_\az n_\bz = - \frac{1}{16
\pi} \sigma^{\az \bz} \pounds_n \sigma_{\az \bz} \nn
\eea
and so is the extrinsic curvature of $\Omega_t$ with respect to a surface
(locally) defined by the tangent vectors of $\Omega_t$ and the
normal vector $n^\az$. 
Geometrically it measures how the area of $\Omega_t$ changes
in the direction $n^\az$. 

For asymptotically flat spacetimes we define the reference 
term $\underline{E}$ so that $E$
will vanish for flat space $\underline{M}$. The simplest way to do this is to (locally) embed $B$ in $\underline{M}$ and then define
$\underline{\tilde{\varepsilon}}$ in the same way as
$\tilde{\varepsilon}$. Specifically we embed
the two surface $\Omega_t$ and then define a vector field
$\underline{T}^\az$ over
$\Omega_t$ such that $\underline{T}^\az \underline{T}_\az = T^\az T_\az$,
$\pounds_{\underline{T}} \underline{\sigma}_{\az \bz} =
\pounds_T \sigma_{\az \bz}$, and $\underline{T}^\az
\underline{\sigma}_\az^\bz = 
T^\az \sigma_\az^\bz$. Then we define a reference quasilocal energy
density as
\bea
\underline{\oep} = \frac{1}{8 \pi} \underline{\tilde{k}} =  
- \frac{1}{8 \pi} \underline{\sigma}^{\az \bz}
\underline{\nabla}_\az \underline{n}_\bz = - \frac{1}{16
\pi} \underline{\sigma}^{\az \bz} \pounds_{\underline{n}}
\underline{\sigma}_{\az \bz}. \nn
\eea
Finally, the matter term is
\bea
\oep^m = -\frac{1}{4\pi} (n_\az \tilde{E^\az})(u^\bz A_\bz) \nn,
\eea
where $\tilde{E}^\az \equiv e^{-2a\phi} F_{\az \bz} \tilde{u}^\bz$ is
the electric field. $\oep^m$,
may be thought of as a charge times a Coulomb potential.
Roughly it represents the potential energy of the region $M$ with
respect to the EM potential $A_\az$. Note that it is gauge dependent.
Three natural gauge choices set $\varepsilon^m = 0$ at infinity,
the quasilocal surface $\Omega_t$, or the black hole horizon. 

In the following we consider two quasilocal energies. The first is
$E_{tot}- \underline{E}$ with the gauge 
chosen so that $\tilde{\varepsilon}^m = 0$ on the black hole horizon, and the second is the 
geometrical energy $E_{Geo} - \underline{E}$ 
where we have chosen the gauge so $\tilde{\varepsilon}^m = 0$ 
on $\Omega_t$. Then, 
\bea
E_{Geo} = \int_{\Omega_t} d^2x \sqrt{\sigma}
\tilde{\varepsilon}, \ \
E_{Tot} = \int_{\Omega_t} d^2x \sqrt{\sigma}
( \tilde{\varepsilon} + \tilde{\varepsilon}^m ),
\ \ \mbox{and} \ \ 
\underline{E} = \int_{\Omega_t} d^2x \sqrt{\sigma}
\underline{\tilde{\varepsilon}}.
\eea

\subsection{Transformation laws for the boosted QLE}

Next we investigate how the QLE transforms with respect to motion of the
observers. 
Consider two sets of observers who instantaneously coincide on a surface
$\Omega_t$. Here we take them as a ``static'' set being evolved
by the timelike unit vector $T^\az = u^\az$ with normal vector
$n^\az$ to $\Omega_t$ and a 
``moving'' set being evolved by $T^{\ast \az} = u^{\ast \az}$.
Then the moving set are seen to have velocity
$ v = -\frac{T^{\ast \az} n_\az}{T^{\ast \az} u_\az}$
in the direction $n^\az$ by the static set. 
Defining $\gamma = (1-v^2)^{-1/2}$, it is not hard to show that
\bea
\varepsilon^\ast = \gamma ( \varepsilon + v j_\vdash) \ \ 
\mbox{and} \ \ 
\varepsilon^{m \ast} = \gamma ( \varepsilon + v j^m_\vdash). \nn
\eea
$j_\vdash = -\frac{1}{16 \pi} \sigma^{\az \bz} \pounds_u \sigma_{ \az
\bz}$ and so represents the (local) rate of change of the area of
$\Omega_t$ as measured by the observers with respect to proper time. It
can be thought of as a momentum flow through the surface. $j^m_\vdash$ 
is a matter term that can be set to zero by an appropriate gauge choice
(which we'll make here for simplicity). Thus, if the $j_\vdash$ terms are
zero, the transformation laws are very similar to those
for energy in special
relativity. 

Things are slightly complicated because the reference terms transform with
respect to a different velocity. Looking back at the defining conditions
for the reference term we see that by construction
$\underline{j}_\vdash = j_\vdash$. Then, it is not surprising that the
surface of observers would have to travel at a different speed in the
reference spacetime than they do in the original one to keep this 
rate of change the same. Thus, 
\bea
\underline{\varepsilon}^\ast = \underline{\gamma} (
\underline{\varepsilon} + \underline{v} j_\vdash) \nn
\eea
where $\underline{v}$ is defined in an analogous way to $v$. 

With these transformation laws it is also easy to see that 
$\varepsilon^{\ast 2} - j_\vdash^{\ast 2}$ is a constant, independent of
the boost. This is analogous to the special relativity relation $E^2 - 
p^2 c^2 = m^2$ (which is a constant) and will be of use in the later
calculations. 

\section{Naked black holes}

Naked black holes are a subclass of the low-energy-limit 
string theory solutions with metric
\bea
ds^2 = -F(r) dt^2 + \frac{dr^2}{F(r)} + R(r)^2 
( d \theta^2 + \sin^2 \theta d \varphi^2 ), \nn
\eea
where $F(r) = \frac{(r-r_+)(r-r_-)}{R^2}$ and
$R(r) = r \left( 1 - \frac{r_-}{r} \right)^{a^2/(1+a^2)}$.
$r_+$ is the location of the black hole horizon and (for the coupling
constant $a \neq 0$) $r_-$ gives the position of the singularity 
behind the horizon.
There are also dilaton and Maxwell fields defined by
\bea
\phi = -\frac{a}{1+a^2} \ln \left(1 - \frac{r_-}{r} \right)
\ \ \mbox{and} \ \  
F= G_0 \sin \theta d\theta \wedge d\varphi.
\eea
The (ADM) mass and magnetic charge of these solutions
are given by
$M = \frac{r_+}{2} + \frac{1-a^2}{1+a^2} \frac{r_-}{2}$ 
and
$G_0 = \left( \frac{r_+ r_-}{1+a^2} \right)^{1/2}$.
If $a=0$  this solution reduces to a
magnetically charged Reissner-Nordst\"{o}m (RN) black hole. For the
purposes of
this short paper 
we shall also assume that if $a \neq 0$ then $a \approx 1$.
Very small values of $a$ cause complications; we deal with these elsewhere
\cite{naked}. 

If $R_+ = R(r_+) \gg 1$ (that is $R_+$ is much larger than the
Planck length) then these black holes have a very large surface
area and a correspondingly large mass (in the corresponding
Planck units). 
All of the curvature invariants are small outside of the horizon, 
and static ($r = \mbox{constant}$, 
$t= \mbox{constant}$ ) observers measure very small 
curvature components. Members of
a spherical set of these observers will naturally carry an orthonormal
tetrad 
$\left\{ 
u^\az = F^{-1/2} \partial_t, 
\hat{\theta}^\az = R^{-1} \partial_\theta,
\tilde{n}^\az = F^{1/2} \partial_r, 
\hat{\phi}^\az = (R \sin \theta)^{-1} \partial_\varphi
\right\}$, where with $N=1$ and $V^\az = 0$, $T^\az = u^\az$
defines the evolution of the observers, $n^\az$ is the normal to 
the spherical surface, and the other two components point along that
surface. Then a typical curvature component measured by a set of these
observers ``hovering'' near to the horizon is
$\cR_{u \hat{\varphi} u \hat{\varphi}} \propto R_+^{-2} \ll 1$.
All other measures of the curvature (including curvature invariants)
measured by such static observers are similarly small.

If, however, these holes are also extremely close to being extreme with 
$\delta \equiv
\left( 1 - r_-/r_+ \right)^{\frac{1}{1+a^2}}
\ll a/R_+, $
then observers who are falling into these holes tell a very different story about the curvature components. Consider a spherical set
of observers who started out 
with velocity zero at some very large $r$ and then fell
towards the black hole along a radial geodesic. Then they
naturally carry a tetrad 
$\left\{
\tilde{u}^\az = \gamma ( u^\az + v \tilde{n}^\az ), 
\hat{\theta}^\az = R^{-1} \partial_\theta,
n^\az = \gamma (\tilde{n}^\az + v u^\az), 
\hat{\phi}^\az = (R \sin \theta)^{-1} \partial_\varphi
\right\},$ where again $\tilde{T}^\az = \tilde{u}^\az$ describes their
evolution and $\hat{\phi}^\az$ and $\hat{\theta}^\az$ point along 
the spherical surfaces. $v = -(1-F)^{1/2}$ is the radial 
velocity of the infalling observers as seen by the
static ones, and $\gamma = (1-v^2)^{-1/2}$ is the standard Lorentz
factor from special relativity.
Then, a typical curvature component seen by
such observers as they cross the black hole horizon is
$ \cR_{\tilde{u} \hat{\varphi} \tilde{u} 
\hat{\varphi}} \propto a^2(R_+ \delta)^{-2} \gg 1$.
That is they see extremely large, Planck scale
curvatures. The resulting huge 
geodesic deviation laterally crushes them.
Horowitz and Ross \cite{HR} 
dubbed this subclass of Maxwell-dilaton holes 
{\it naked} because Planck scale curvature components 
could be seen outside of their horizons. 

\subsection{QLE of naked black holes}

These black holes seem almost tailor-made to be investigated by our method
of defining boosted quasilocal energies. A
static  $(r=\mbox{constant},
t=\mbox{constant})$ set of observers measure 
\bea
E_{Geo} - \underline{E} = R - \sqrt{(r-r_+)(r-r_-)} \dot{R}
\ \ \mbox{
   and} \ \
E_{Tot} - \underline{E} = R \left( 1 - \sqrt{\frac{r-r_+}{r-r_-}}
\right), \nn
\eea 
where $\dot{R} = \frac{dR}{dr}$. 
Note that both of these go to $R_+ \gg 1$ at the horizon. This is not
surprising since both measure the quasilocal energy
and large $R_+$ corresponds to a large $M$ black hole.
Keep in mind that for a near extreme magnetic RN hole 
$R_+ \approx 2m$. 

Next consider the infalling measurements. The radial velocity of the
infalling observers in the original spacetime is $v = -(1-F)^{1/2}$ 
with respect to the static observers. By contrast, the shell of observers have to travel at $\underline{v} = 
- \dot{R} (1-F)^{1/2} / ( 1 + \dot{R}^2 (1-F) )^{1/2}$ in the reference
spacetime if they want their surface area to change at the same rate. 
Therefore at the horizon $r_+$,
\bea
E^\ast_{Geo} - \underline{E}^\ast = 
\left\{
\begin{array}{ll}
C_1 R_+ \delta \ll 1 & \mbox{for naked black holes} \\
C_2 R_+ \gg 1        & \mbox{for ``clothed'' holes}
\end{array}
\right\} \ \ \mbox{and} \ \ 
E^\ast_{Tot} - \underline{E}^\ast = \frac{R_+}{\delta} \gg \gg 1,
\eea
where $C_1$ and $C_2$ are constants that are on the order of unity.
Thus we see that our extension of the QLE formalism detects the difference
between regular ``clothed'' and naked black holes.
Note however that while static/infalling observers see small/large
curvatures they measure large/small geometric QLE's.

\subsection{Why do naked black holes behave this way?}

These small/large measurements can be understood physically in the
following manner. As was noted by Horowitz and Ross for a
naked black hole $R_+ \delta$ is more-or-less the time left to an
infalling observer before she reaches the singularity at $r_-$.
Less rigorously, for near extreme black holes $r_+ \approx r_-$ and so in
some sense for naked black holes the singularity is ``just behind'' the
horizon. 

At the singularity the surface area of an $r=\mbox{constant}$ shell of
observers goes to zero. Intuitively this means that for the naked black
holes we expect the magnitude of $j_\vdash$ (basically the rate of change
of the
area) to be very large since the area is very large but will soon be
zero. By contrast for an RN black hole, the area only goes to zero
at $r=0$ which is not so ``close'' to the horizon. Thus $j_\vdash$ need
not be so large. These expectations are borne out by the
calculations.

Thus, we can see why observers falling into a naked black hole experience
the huge lateral crushing forces. As a shell of them travelling on
geodesics cross the horizon the surface area of that shell is rapidly
decreasing and so the crushing lateral forces are to be expected. By
contrast
for ``clothed'' black holes the rate of change of the area is much smaller
and so are the corresponding lateral forces.  

The relative sizes of $j_\vdash$ in the two cases also explain the
geometrical QLE observations. As we saw earlier, 
$\varepsilon^2 - j^2_\vdash$ is a constant independent of boosts.
Therefore if $\varepsilon$ and  $\underline{\varepsilon}$ are the
geometric and reference QLE densites for the static
observers, 
$\varepsilon^\ast$ and  $\underline{\varepsilon}^\ast$ are their boosted
counterparts, and recalling that $\underline{j_\vdash} = j_\vdash $, 
\bea
\varepsilon^\ast - \underline{\varepsilon}^\ast = 
\sqrt{j_\vdash^2 + \varepsilon^2} - \sqrt{j_\vdash^2 +
\underline{\varepsilon}^2} \approx \frac{\varepsilon^2 -
\underline{\varepsilon}^2}{2 j_\vdash}
\eea
for $j_\vdash$ much larger than $\varepsilon$ and $\varepsilon^\ast$. Thus
as $j_\vdash$ becomes larger and larger, $\varepsilon^\ast -
\underline{\varepsilon}^\ast$ becomes smaller and smaller. Physically,
though $\varepsilon^\ast$ and $\underline{\varepsilon}^\ast$ are boosted
to be very large, at the same time the difference between them becomes
smaller and smaller. 

By contrast $E_{Tot}^\ast - \underline{E}^\ast$ includes matter terms.
These terms are also boosted to be very large but there is no
corresponding term in the reference spacetime to cancel them out (as there
is for the geometrical energy). Therefore in $E_{Tot}^\ast -
\underline{E}^\ast$ the matter terms dominate over the geometrical ones
which are small and so the total infalling quasilocal energy is large. 

\section*{Acknowledgements}
This work was supported by the Natural Sciences and Engineering Council of Canada.

\end{document}